\documentclass{iau}
\usepackage{graphicx} 

\title[IAUS290.~~Connecting accreting X-ray and radio MSPs] 
{On the connection between accreting X-ray and radio millisecond pulsars}

\author[T.M.~Tauris]  
{T.M.~Tauris$^{1,2,*}$
 }

\affiliation{$^1\,$Argelander-Institut f\"ur Astronomie, Universit\"at Bonn, Germany\\ 
             $^2\,$Max-Planck Institut f\"ur Radioastronomie, Bonn, Germany \\ 
             $^{*}\,$email: {\tt tauris@astro.uni-bonn.de} \\[\affilskip]
}

\pubyear{2012}
\volume{290}  
\jname{\mbox{Feeding Compact Objects: Accretion on All Scales}}
\editors{C.M. Zhang, ed.} 
\begin{document}

\maketitle

\begin{abstract}
For many years it has been recognized that the terminal stages of 
mass transfer in a low-mass X-ray binary (LMXB) should cause the magnetosphere
of the accreting neutron star to expand, leading to a braking torque acting on
the spinning pulsar.
After the discovery of radio millisecond pulsars (MSPs) it was therefore 
somewhat a paradox \cite[(e.g. Ruderman~et~al.~1989)]{rst89} 
how these pulsars could retain their fast spins 
following the Roche-lobe decoupling phase, RLDP.
Here I present a solution to this 
so-called ``turn-off problem'' which was recently found 
by combining binary stellar evolution models with 
torque computations  \cite[(Tauris~2012)]{tau12}.
The solution is that during the RLDP the spin equilibrium of the pulsar is broken
and therefore it remains a fast spinning object. I briefly discuss these
findings in view of the two observed spin distributions in the populations of 
accreting X-ray millisecond pulsars (AXMSPs) and radio MSPs.

\end{abstract}

\firstsection 
\section{Introduction}
Theoretical modelling of accretion torques with 
disk-magnetosphere interactions have been performed for nearly four decades \cite[(e.g. Ghosh \& Lamb 1979)]{gl79}. 
Likewise, detailed calculations of binary evolution have demonstrated that both the duration of the RLO in LMXBs and
the amounts of matter transfered are adequate to fully 
recycle a pulsar \cite[(e.g. Tauris \& Savonije~1999]{ts99}; \cite[Podsiadlowski~et~al.~2002)]{prp02}.
However, these previous studies did not combine numerical stellar
evolution calculations with computations of the resulting accretion torque at work.
For further details on these issues, and more general discussions of my results, I refer to 
the journal paper, \cite[Tauris~(2012)]..

\section{The magnetosphere-disk interactions}
The interplay between the neutron star magnetic field and the conducting plasma in the accretion disk
is a rather complex process.
The physics of the transition zone from Keplerian disk to magnetospheric flow is important and
determines the angular momentum exchange from the differential rotation between the disk
and the neutron star \cite[(e.g. Spruit \& Taam~1993)]{st93}.
It is the mass transfered from the donor star which carries this angular momentum which eventually
spins up the rotating neutron star once its surface magnetic flux density, $B$
is low enough to allow for efficient accretion.

The gain in neutron star spin angular momentum can approximately be expressed as:
$\Delta J_\star = \sqrt{GMr_{\rm mag}}\,\Delta M \,\xi$,
where $\xi \simeq 1$ is a numerical factor which depends on the flow pattern \cite[(Ghosh \& Lamb~1979)]{gl79},
$\Delta M = \dot{M}\cdot\Delta t$ is the amount of mass accreted in a time interval $\Delta t$
with average mass accretion rate $\dot{M}$ and $r_{\rm mag}$ is the magnetospheric boundary, roughly located at 
the inner edge of the disk; i.e. 
$r_{\rm mag}=\phi\cdot r_{\rm Alfven}$, where the magnetospheric coupling parameter, 
$\phi = 0.5-1.4$, \cite[Wang~(1997)]{wan97}; \cite[D'Angelo \& Spruit~(2010)]{ds10}. 
For the numerical calculations of the RLDP, I included the effect of additional spin-down torques, acting on the neutron star,
due to both magnetic field drag on the accretion disk \cite[(Rapapport~et~al.~2004)]{rfs04} as well as magnetic dipole radiation,
although these effects are usually not dominant.
The total spin torque can be written as:
\begin{equation} 
\label{eq:drag}
  N_{\rm total} = n(\omega)\left(\dot{M}\sqrt{GMr_{\rm mag}}\,\xi + \frac{\mu ^2}{9r_{\rm mag}^3}\right) - \frac{\dot{E}_{\rm dipole}}{\Omega}
\end{equation}
where
$n(\omega)=\tanh \left( (1-\omega)/\delta _\omega \right)$
is a dimensionless function, depending on the fastness parameter,
$\omega = \Omega _\star /\Omega_{\rm K}(r_{\rm mag}) = (r_{\rm mag}/r_{\rm co})^{3/2}$, which is introduced
to model a gradual torque change in a transition zone near the magnetospheric boundary.

\section{Roche-lobe decoupling phase (RLDP)}
\begin{figure}[b]
\begin{center}
 \includegraphics[width=3.2in, angle=-90]{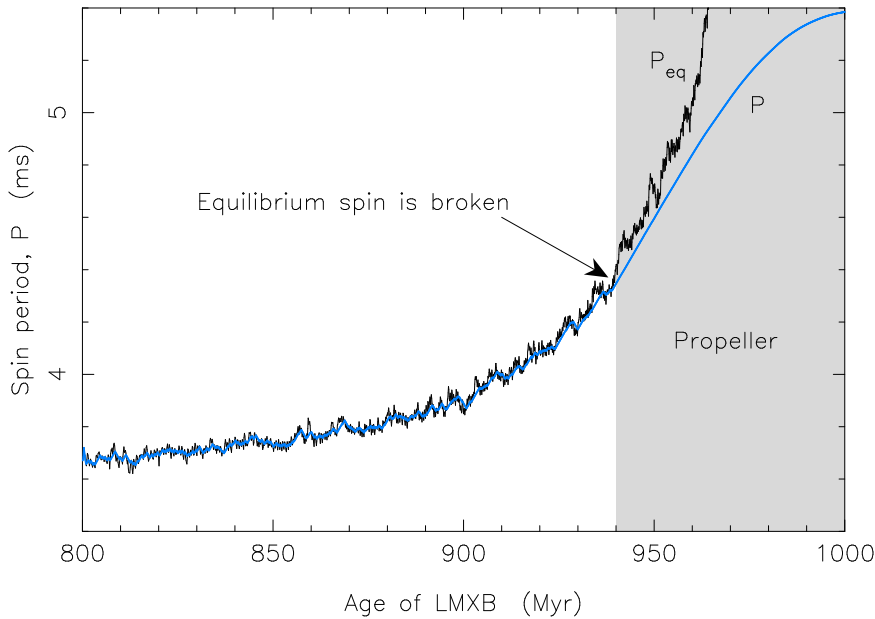} 
 \caption{Transition from equilibrium spin to propeller phase. 
   The black line is the equilibrium spin period of the neutron star, $P_{\rm eq}$ (the oscillations reflect
   fluctuations in $\dot{M}(t)$) and the blue line is its actual spin period, $P$.
   At early stages
   of the Roche-lobe decoupling phase (RLDP) the neutron star spin is able to remain
   in equilibrium despite the outward moving magnetospheric boundary caused by
   decreasing ram pressure. However, at a certain point (indicated by the arrow), when
   the mass-transfer rate decreases rapidly, the torque can no longer transmit the
   deceleration fast enough for the neutron star to remain in equilibrium. 
   (\cite[Tauris~2012]{tau12}.)
} 
\label{fig1}
\end{center}
\end{figure}
\begin{figure}[b]
\begin{center}
 \includegraphics[width=3.8in, angle=-90]{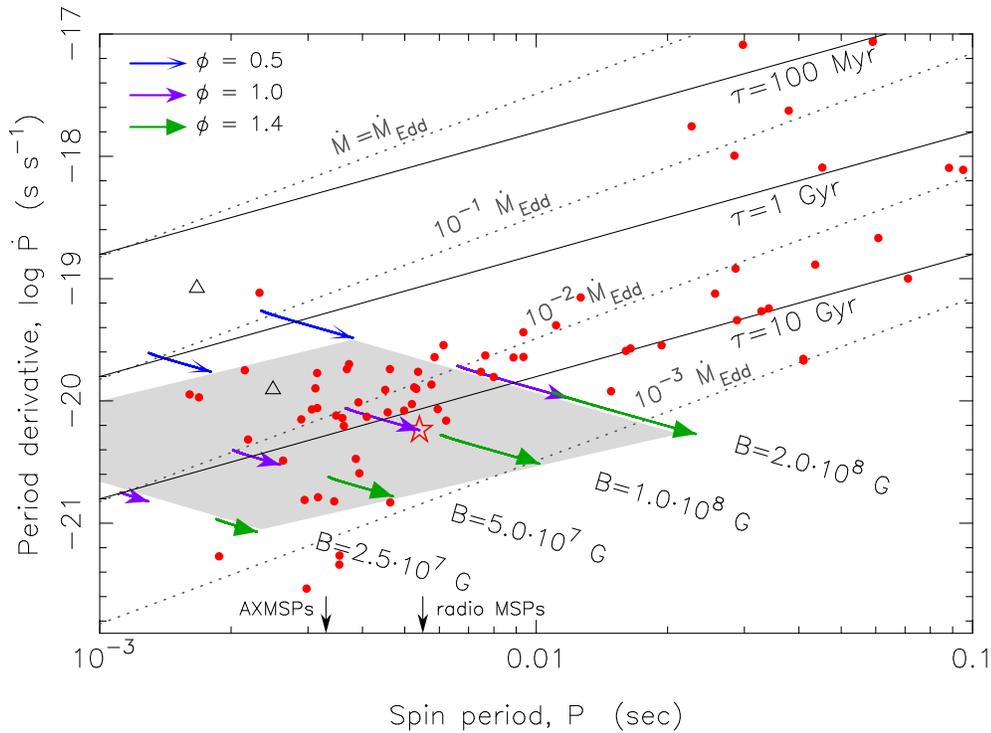} 
 \caption{Evolutionary tracks during the Roche-lobe decoupling phase (RLDP).
   Computed tracks are shown as arrows in the $P\dot{P}$--−diagram calculated by using different
   values of the neutron star B-field strength. The various types of arrows correspond to
   different values of the magnetospheric coupling parameter, $\phi$. The gray-shaded area
   indicates all possible birth locations of recycled MSPs calculated from one donor star
   model (marked by a star). The solid lines represent characteristic ages, $\tau$ , and the dotted lines are spin-up
   lines calculated for a magnetic inclination angle, $\alpha = 90^{\circ}$. 
   The two triangles indicate approximate
   locations of the AXMSPs SWIFT J1756.9−2508 (upper) and SAX 1808.4−3658
   (lower). Observed MSPs in the Galactic field are shown as dots [data taken from the ATNF
   Pulsar Catalogue, December 2011]. All the measured $\dot{P}$ values are corrected for the
   Shklovskii effect.
   The average spin periods
   of AXMSPs and radio MSPs are indicated with arrows at the bottom of the diagram.
   (\cite[Tauris~2012]{tau12}.)
} 
\label{fig2}
\end{center}
\end{figure}

The rapidly decreasing mass-transfer rate during the RLDP
results in an expanding magnetosphere which causes a significant braking torque to act on the spinning pulsar.
Thus it forces the rotational period to increase, as shown in Fig.~\ref{fig1}. 
At some point the spin equilibrium is broken. Initially, the spin can remain in equilibrium by adapting to the decreasing
value of $\dot{M}$. Further into the RLDP, however, $r_{\rm mag}$ increases on a timescale faster than
the spin-relaxation timescale, $t_{\rm torque}$ at which the torque can transmit the effect of deceleration to
the neutron star and therefore $r_{\rm mag} > r_{\rm co}$, leading to $P < P_{\rm eq}$ at all times onwards.
The results is that AXMSPs in LMXBs typically lose $\sim\!50$~\% of their
rotational energy (see Fig.~\ref{fig2}), depending on the B-field of the pulsar, the duration of the RLDP and
the assumptions governing the disk-magnetosphere interactions.

\section{The spin-relaxation timescale and the duration of the RLDP}
To estimate the spin-relaxation timescale one can simply consider: $t_{\rm torque}=J/N$:
\begin{equation} 
\label{eq:timescale}
    t_{\rm torque} = I \left(\frac{4G^2M^2}{B^8R^{24}\dot{M}^3}\right) ^{1/7} \frac{\omega _c}{\phi^2\,\xi} 
  \;\simeq \quad 50\,{\rm Myr}\quad B_8^{-8/7}\left(\frac{\dot{M}}{0.1\,\dot{M}_{\rm Edd}}\right) ^{-3/7} \left(\frac{M}{1.4\,M_{\odot}}\right) ^{17/7}
\end{equation}
In intermediate-mass X-ray binaries (IMXBs) 
the mass-transfer phase from a more massive companion is relatively short, 
causing the RLDP~effect to be negligible.
The reason for the major difference between the RLDP~effect in LMXBs (often leading to the 
formation of radio MSPs with helium white dwarf companions) and IMXBs (primarily leading to mildly recycled pulsars
with carbon-oxygen/oxygen-neon magnesium white dwarfs) is the
time duration of the RLDP relative to the spin-relaxation timescale, i.e. the ratio of $t_{\rm RLDP}/t_{\rm torque}$.
In LMXBs, $t_{\rm RLDP}/t_{\rm torque}\simeq\!0.3$. However, in IMXBs this ratio is often smaller by a factor of 10
which causes the spin period to ``freeze'' at the original value of $P_{\rm eq}$,
see \cite[Tauris~et~al.~(2012) for further details and examples of model calculations]{tlk12}.

\section{Spin distribution of AXMSPs and radio MSPs}
\begin{figure}[t]
\begin{center}
 \includegraphics[width=2.3in, angle=-90]{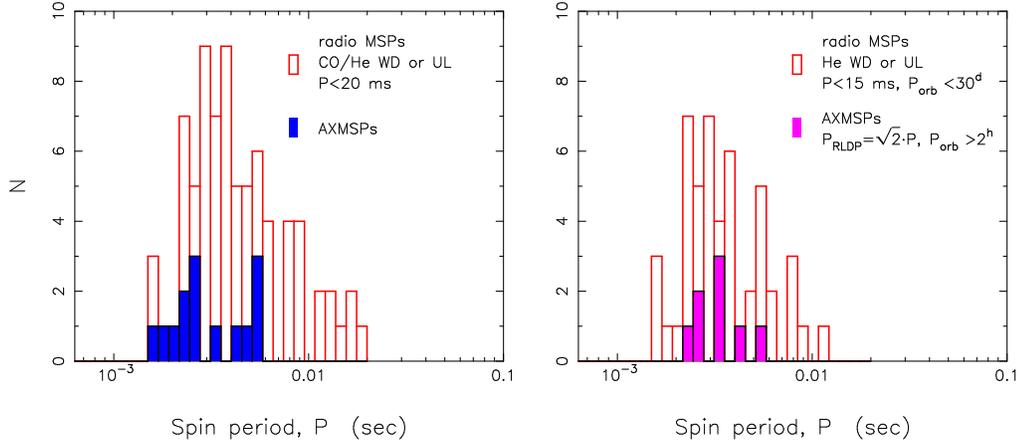} 
 \caption{Left panel: the observed distribution of all 78 Galactic radio MSPs (open red bins) 
          and 14 AXMPS (solid blue bins).
          Right panel: 49 radio MSPs with He~WD or ultra-light companions, $P<15\;{\rm ms}$ and $P_{\rm orb}<30^{\rm d}$,
          together with the
          8 AXMSPs, with $P_{\rm orb}>2^{\rm h}$ which are likely to evolve into binary radio MSPs, corrected for the
          RLDP~effect when these sources become radio MSPs by multiplying their spin periods by a factor of $\sqrt{2}$
          (solid pink bins). 
          The good correspondence between these two populations 
          supports the RLDP-effect hypothesis.
           } 
   \label{fig3}
\vspace{1mm}
\end{center}
\end{figure}
The RLDP~effect discussed here can explain why the recycled radio MSPs (observed {\it after} the RLDP)
are significantly slower rotators compared to
the more rapidly spinning AXMSPs (observed {\it before} the RLDP), see Fig.~\ref{fig3}.
Only for radio MSPs with $B> 10^8\,{\rm G}$ can the difference in spin periods be partly understood from regular magnetic dipole
and plasma current spin-down over a radio MSP lifetime of several Gyr.
When comparing radio MSPs and AXMSPs one should be aware of differences in their
binary properties and observational biases, see SOM in \cite[Tauris~(2012)]{tau12}. 
Observationally, the RLDP~effect can be verified
if future surveys discover a significant number of AXMSPs and radio MSPs confined to
an interval with similar orbital periods and companion star masses.

Although AXMSPs are believed to be progenitors of  radio MSPs,
all of the 14 observed AXMSPs have orbital periods less than one day whereas fully recycled radio MSPs
are observed with orbital periods all the way up to a few hundred days. This is a puzzle.
Some radio MSPs are born (recycled) with $B\simeq 1\times 10^7\,{\rm G}$ 
and they would most likely not be
able to channel the accreted matter sufficiently to become
observable as AXMSPs.

Future modelling of the RLDP should ideally include irradiation effects, 
accretion disk instabilities and cyclic accretion (e.g. \cite[Spruit \& Taam~1993]{st93}; \cite[D'Angelo \& Spruit~2010]{ds10}).

\end{document}